%% file: main.tex
\documentclass{webofc}
\usepackage[varg]{txfonts}   
\usepackage{hyperref}
\usepackage{lineno}
\usepackage{lipsum}
\input{commands}

\begin{document}
%
\title{Small systems at the LHC}
%
%

\author{
  \firstname{Roberto} \lastname{Preghenella}\inst{1}\fnsep\thanks{on behalf of ALICE, ATLAS and CMS Collaborations (\email{preghenella@bo.infn.it})}}

\institute{Istituto Nazionale di Fisica Nucleare, Sezione di Bologna}

\abstract{%
  In these proceedings, I report on a selection of recent LHC results in small systems from ALICE~\cite{Aamodt:2008zz}, ATLAS~\cite{Aad:2008zzm} and CMS~\cite{Chatrchyan:2008aa} experiments. Due to the fact that the investigation of QCD in small systems at high multiplicity is becoming an increasingly large subject, interesting the heavy-ion community and more in general the high-energy physics community, not all the related topics can be discussed in this paper. The focus will be given to some of the measurements addressing the physics of collective phenomena in small systems and to the recent results on strangeness enhancement in proton-proton collisions. The reader must be informed that a large number of interesting results did not find space in the discussion reported here.
}
\maketitle
\input{introduction}
\input{collective}

\input{strangeness}

\input{conclusions}
\bibliography{biblio}
\end{document}

%% file: commands.tex
\usepackage{xspace}
\usepackage{color}
\usepackage{xcolor}
\usepackage{rotating}
\definecolor{light-gray}{gray}{0.8}

\newcommand{\cmnt}[1]{}

\newcommand{\pt}{\ensuremath{p_{\rm T}}\xspace}

\newcommand{\dnchdeta}{\ensuremath{{\rm d}N_{\rm ch}/{\rm d}\eta}\xspace}

\newcommand{\GeVc}{GeV/$c$\xspace}

\newcommand{\pPb}{p--Pb\xspace}
\newcommand{\PbPb}{Pb--Pb\xspace}

\newcommand{\Tkin}{\ensuremath{T_{\rm kin}}\xspace}
\newcommand{\betaT}{\ensuremath{\beta_{\rm T}}\xspace}

\newcommand{\sPi}{\ensuremath{{\pi}}\xspace}

\newcommand{\sProton}{\ensuremath{\rm p}\xspace}

\newcommand{\sPr}{\ensuremath{\rm p}\xspace}

\newcommand{\pKzero}{\ensuremath{{\rm K}^{0}_{S}}\xspace}
\newcommand{\sLambda}{\ensuremath{\Lambda}\xspace}

\newcommand{\sXi}{\ensuremath{\Xi}\xspace}

\newcommand{\sOmega}{\ensuremath{\Omega}\xspace}

%% file: introduction.tex
\section{Introduction}\label{sec:introduction}

There was a time when small collision systems, like proton-proton and proton-nucleus, where almost excusively employed in heavy-ion physics as a reference to compute the nuclear modification factors of particle production and as a proxy to disentangle initial state, cold nuclear matter effects (i.e. due to the nuclear PDF of the target/projectile) from genuine final state, hot nuclear matter effects. As a matter of fact, results based on the analysis of particle production in small systems provided the evidence that the characteristic phenomena observed in heavy-ion collisions, most notably the high-\pt and jet suppression~\cite{Adler:2003ii,Adams:2003im}, were actually due to the formation of a hot state of QCD matter, the Quark-Gluon Plasma.

With the advent of the LHC, pioneering studies of pp collisions in high-multiplicity events revealed the presence of unexpected phenomena in small systems. The discovery of the ``ridge'' in pp collisions by the CMS Collaboration was the first of a long list of such results~\cite{Khachatryan:2010gv}. Long-range, near-side angular correlations in particle production emerged in pp and subsequently in \pPb collisions~\cite{CMS:2012qk} and paved the way for a systematic investigation of the existence of collective phenomena, known since long in heavy-ion collisions~\cite{Abelev:2009af}, in the much smaller pp and \pPb collision systems. A wealth of new, unexpected phenomena have been observed so far with striking similarities to heavy-ion phenomenolgy. In this report, only a few of them can be discussed because of the lack of space. The focus will be given to a selection of recent LHC results aiming at studying collective phenomena and the chemistry of hadron production in small systems.

%% file: collective.tex
\section{Collective phenomena}\label{sec:collective}
The observation of the ridge in pp~\cite{Khachatryan:2010gv} and subsequently in \pPb collisions~\cite{CMS:2012qk} promptly triggered further investigations. Studies of two-particle correlations in \pPb collisions reported by the ALICE Collaboration have shown that, when taking the difference between the yields per trigger particle in high-multiplicity and low-multiplicity events, two nearly identical long-range ridge-like excess structures on the near-side and away-side, the so-called ``double ridge'', arise~\cite{Abelev:2012ola,Aad:2012gla}. Results on two-particle correlations of identified hadrons~\cite{Khachatryan:2016txc,ABELEV:2013wsa,Khachatryan:2014jra,Abelev:2014pua} and of Bose-Einstein (HBT) correlations~\cite{ATLAS:2017jts,Adamova:2017opl} further stressed the initial observations, showing striking similarities between pp, \pPb and \PbPb collisions, consistent with the hydrodynamic picture of a particle-production source expanding more explosively along the collision event plane.

\input{multiparticle}

\input{spectra}

%% file: multiparticle.tex
\begin{figure}[t]
  \centering
  \includegraphics[width=0.9\textwidth]{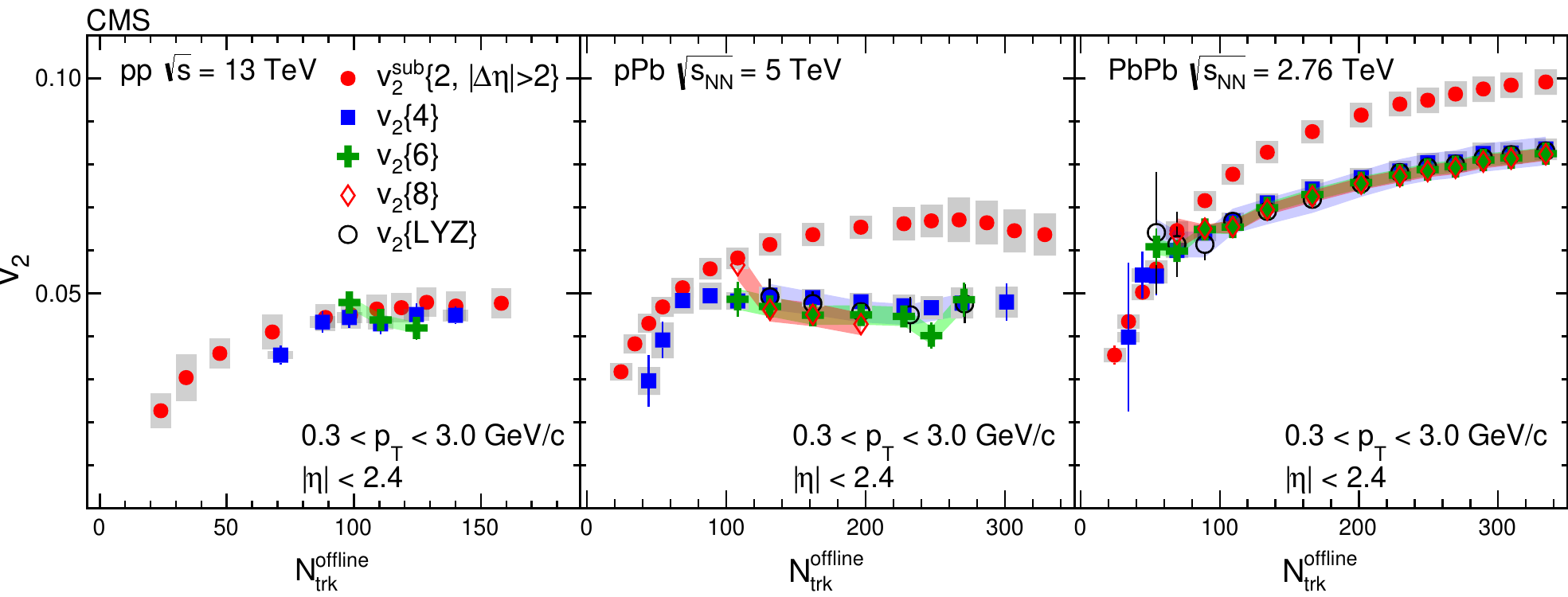}
  \caption{Second-order azimuthal anisotropy Fourier harmonics $v_{2}$ measured by CMS in pp, \pPb and \PbPb collisions based on correlations among multiple charged particles~\cite{Khachatryan:2015waa,Khachatryan:2016txc}.}
  \label{fig:collective:multiparticle:cms}
\end{figure}
\begin{figure}[t]
  \centering
  \includegraphics[width=0.45\textwidth]{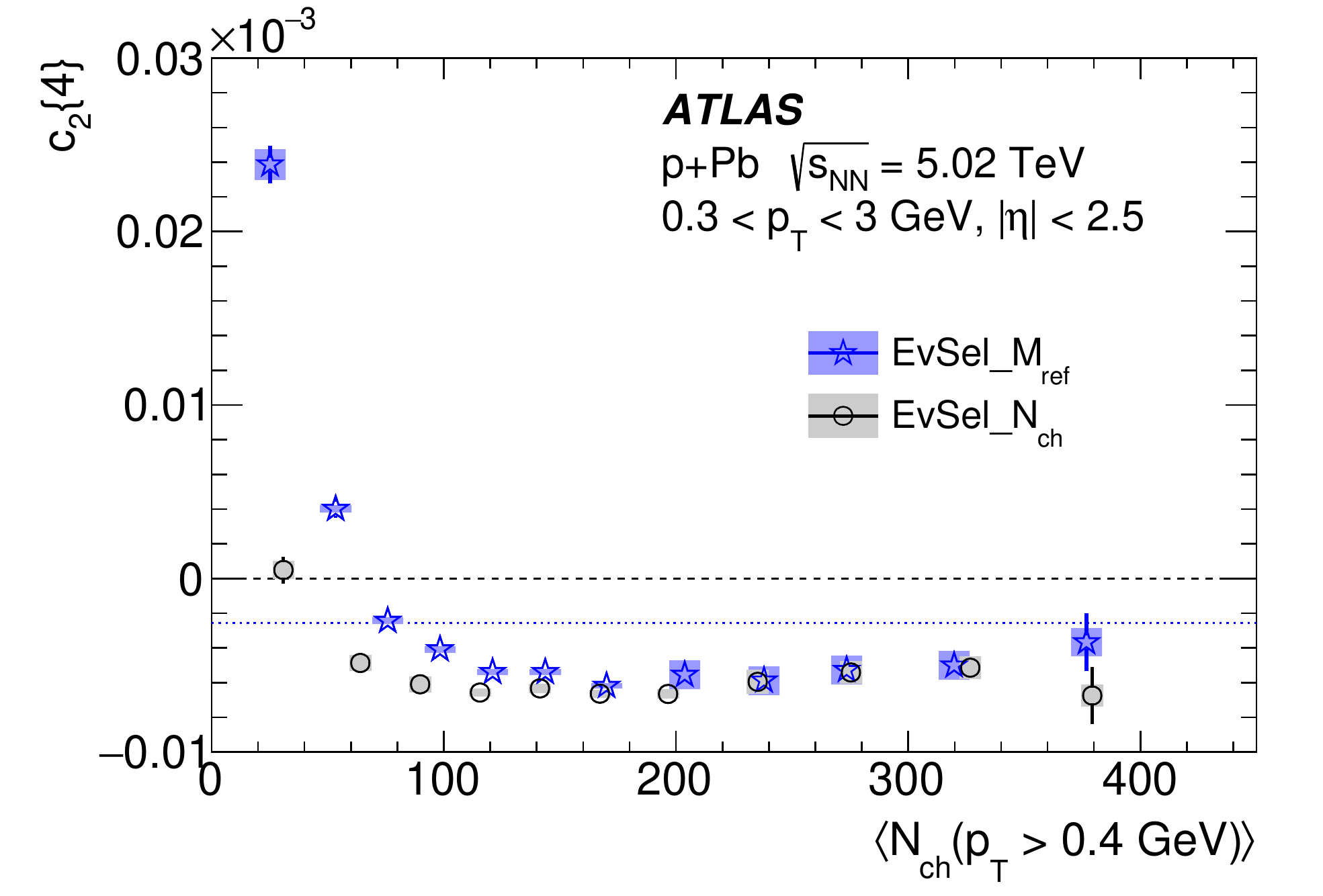}
  \includegraphics[width=0.45\textwidth]{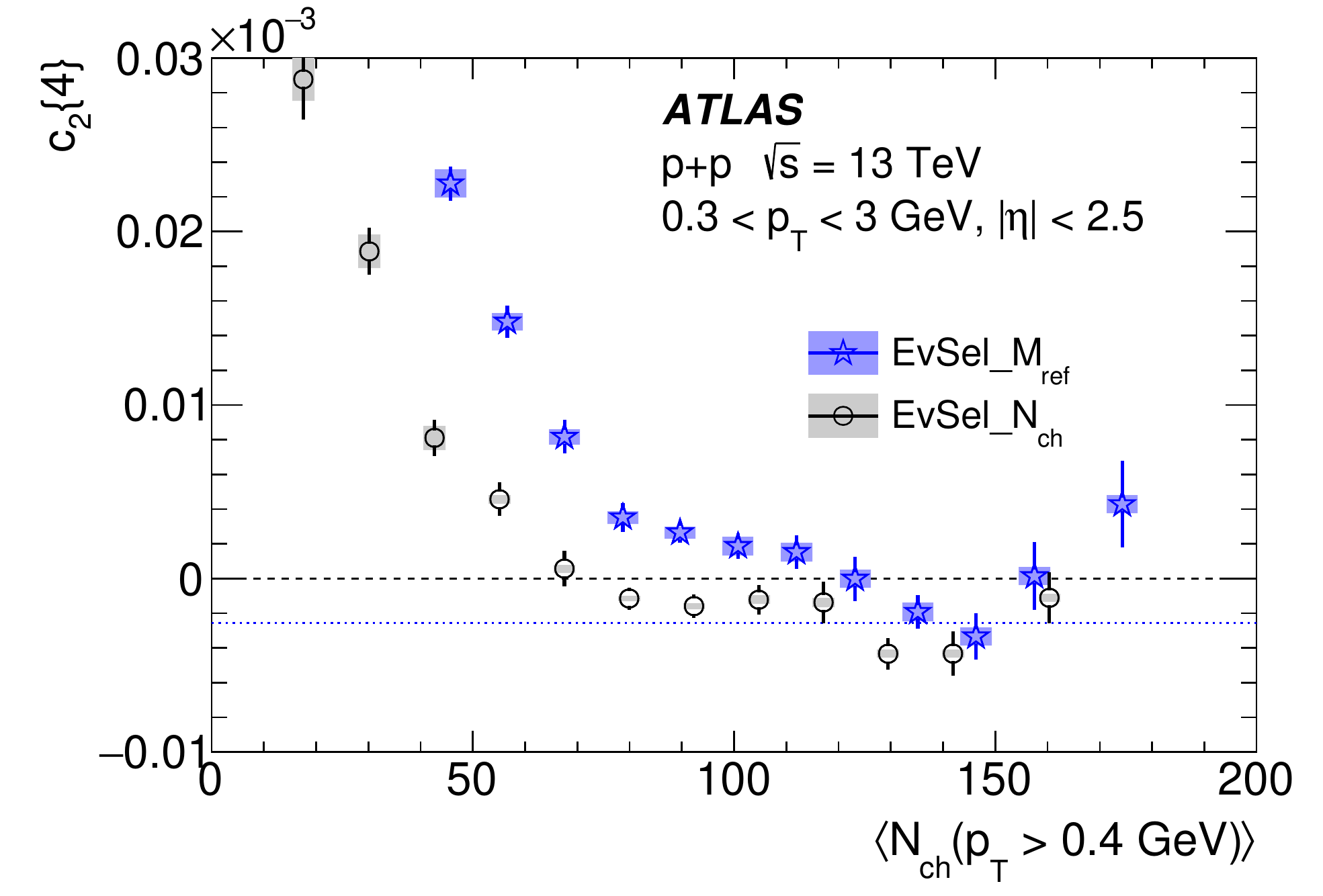}
  \caption{Four-particle cumulants $c_{2}\{4\}$ measured by ATLAS in \pPb collisions (left) and in pp collisions (right) using different multiplicity selection methods~\cite{Aaboud:2017acw}.}
  \label{fig:collective:multiparticle:atlas}
\end{figure}
\begin{figure}[t]
  \centering
  \includegraphics[width=0.45\textwidth]{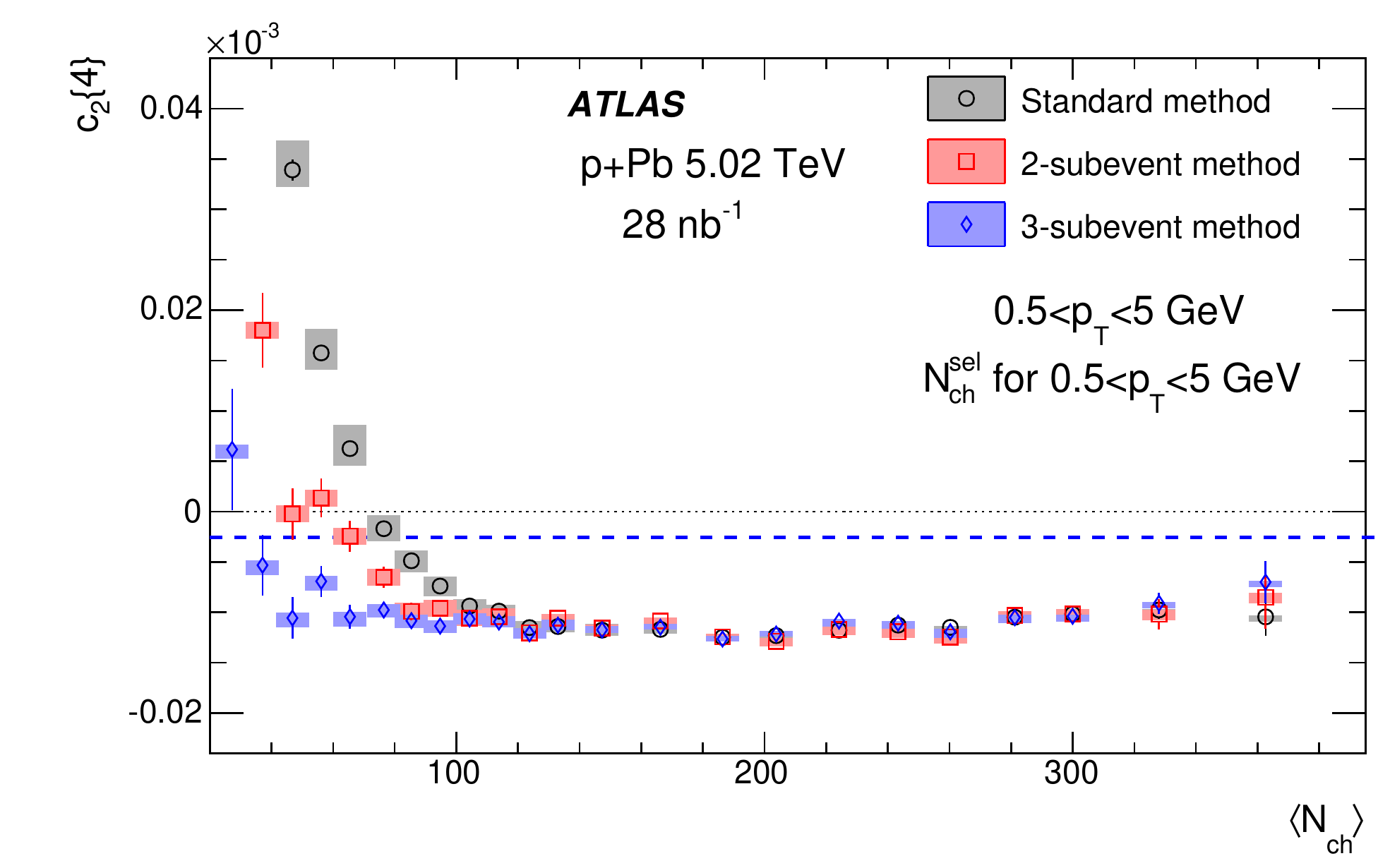}
  \includegraphics[width=0.45\textwidth]{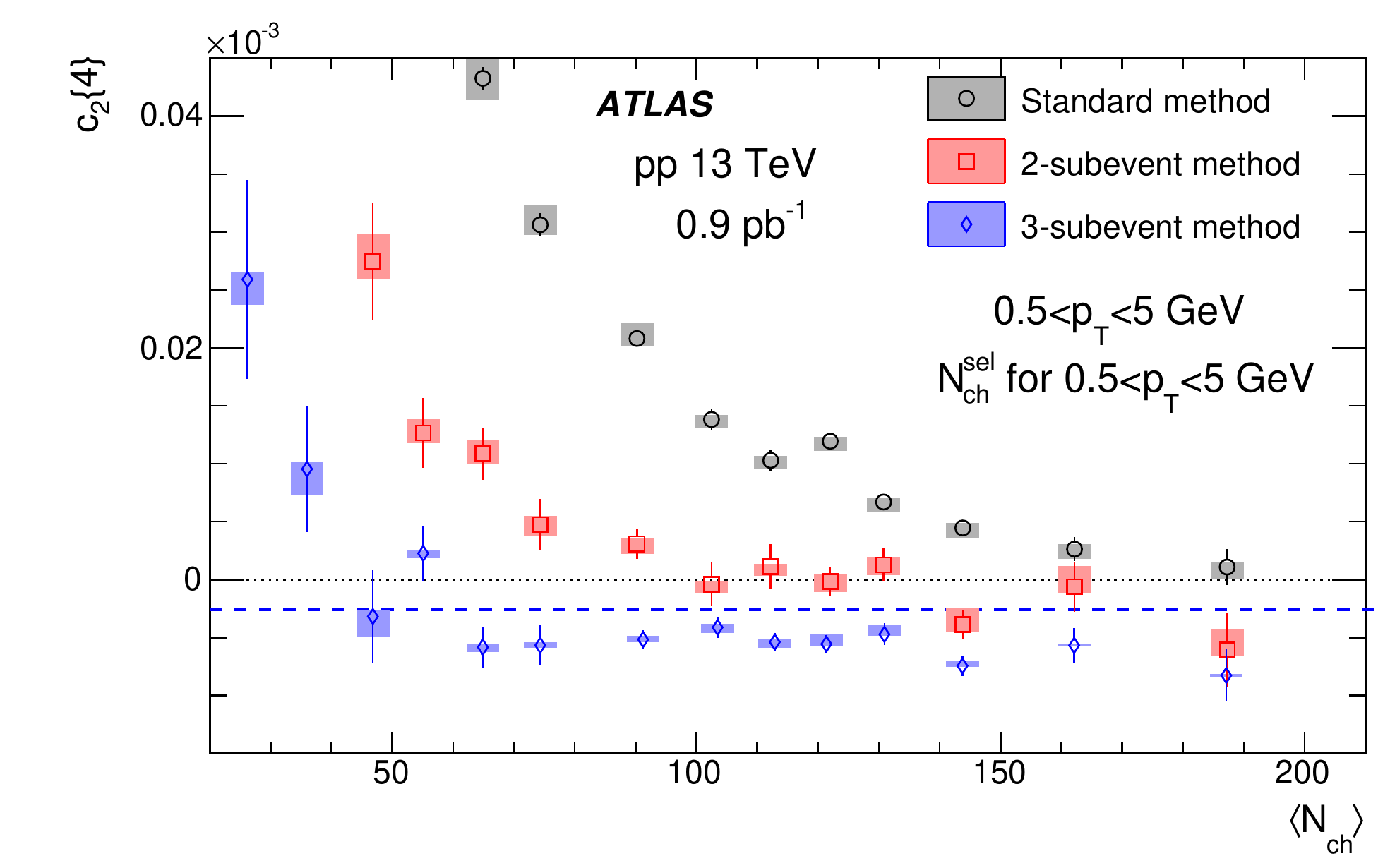}
  \caption{Four-particle cumulants $c_{2}\{4\}$ measured by ATLAS in \pPb collisions (left) and in pp collisions (right) with the ``standard'', ``two-subevent'' and the ``three-subevent'' methods~\cite{Aaboud:2017blb}.}
  \label{fig:collective:multiparticle:atlas2}
\end{figure}

\subsection{Multi-particle correlations}\label{sec:collective:multiparticle}

A fundamental question is whether the two-particle azimuthal correlation structures observed at large relative pseudorapidity in pp and \pPb collisions result from correlations exclusively between particle pairs, or if it is a multi-particle, genuine collective effect. A strong hint, originally named as evidence by the CMS Collaboration, for collective multi-particle correlations in pp and \pPb pp collisions was reported already~\cite{Khachatryan:2015waa,Khachatryan:2016txc}.
Figure~\ref{fig:collective:multiparticle:cms} shows the second-order azimuthal anisotropy Fourier harmonics $v_{2}$ measured in pp, \pPb and \PbPb collisions over a wide pseudorapidity range based on correlations among up to eight particles. The $v_{2}$ values obtained with correlations among more than four particles are consistent with four-particle results and with comparable magnitude to those from two-particle correlations. These observations support the interpretation of a collective origin for the observed long-range correlations in high-multiplicity pp and \pPb collisions.

There is a note of care that must be taken seriously into account. While multi-particle correlation measurements have the advantage of suppressing short-range two-particle correlations such as jets and resonance decays, they are not totally insensitive to such or other so-called ``non-flow'' effects. This has been stressed in a publication by the ATLAS Collaboration~\cite{Aaboud:2017acw}, where it is shown how different multiplicity selection methods may yield different four-particle cumulants ($c_{2}\{4\}$) results. 
The measurements of $c_{2}\{4\}$ are shown in Figure~\ref{fig:collective:multiparticle:atlas} for pp and \pPb collisions. The results confirm the evidence for collective phenomena in \pPb and low-multiplicity \PbPb collisions (not reported in Figure~\ref{fig:collective:multiparticle:atlas}). On the other hand, the pp results for $c_{2}\{4\}$ do not demonstrate collective behaviour, indicating that they may be biased by contributions from non-flow correlations.

Reliably suppressing non-flow correlations in pp collision is a central issue that needs to be solved in order to be able to unveil the actual underlying collectivity. Fortunately, several new methods have been recently developed which aim at tackling this issue. As an example, the ATLAS Collaboration has reported on the measurement of multi-particle correlations in pp and \pPb collisions with the so-called ``two-subevent'' and the ``three-subevent'' methods~\cite{ATLAS:2017tqk}. Figure~\ref{fig:collective:multiparticle:atlas2} shows $c_{2}\{4\}$ measured in pp and \pPb collisions with such methods.  $c_{2}\{4\}$ from the standard method is sensitive to the choice of particles used to form the event classes. The sensitivity is greatly reduced in the two-subevent method and is almost fully removed in the three-subevent method, suggesting that the three-subevent method is much more robust against non-flow effects. The three-subevent method shows significant flow in pp collisions in a broad $N_{\rm ch}$ range.

%% file: spectra.tex
\begin{figure}[t]
  \centering
  \includegraphics[width=0.9\textwidth]{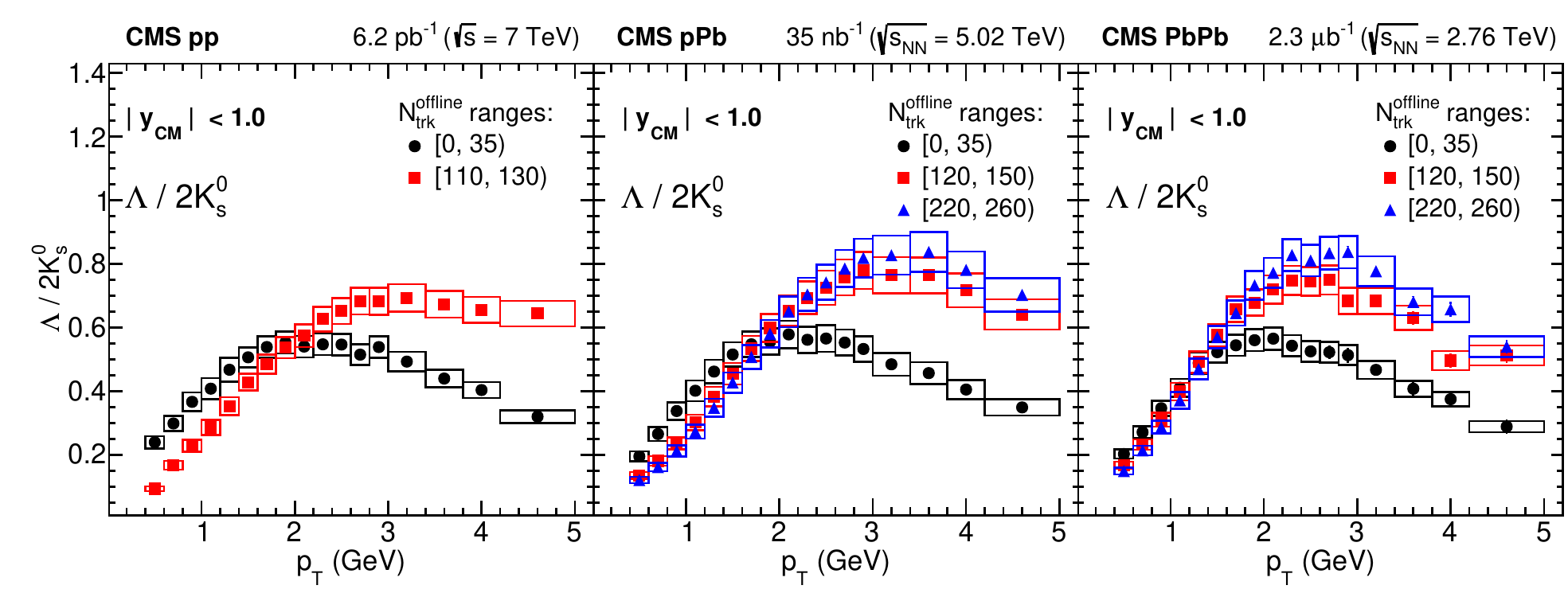}
  \caption{Ratios of \pt spectra of \sLambda/2\pKzero measured by CMS in pp (left), \pPb (middle), and \PbPb collisions (right)~\cite{Khachatryan:2016yru}.}
  \label{fig:collective:spectra:cms}
  \centering
  \includegraphics[width=0.45\textwidth]{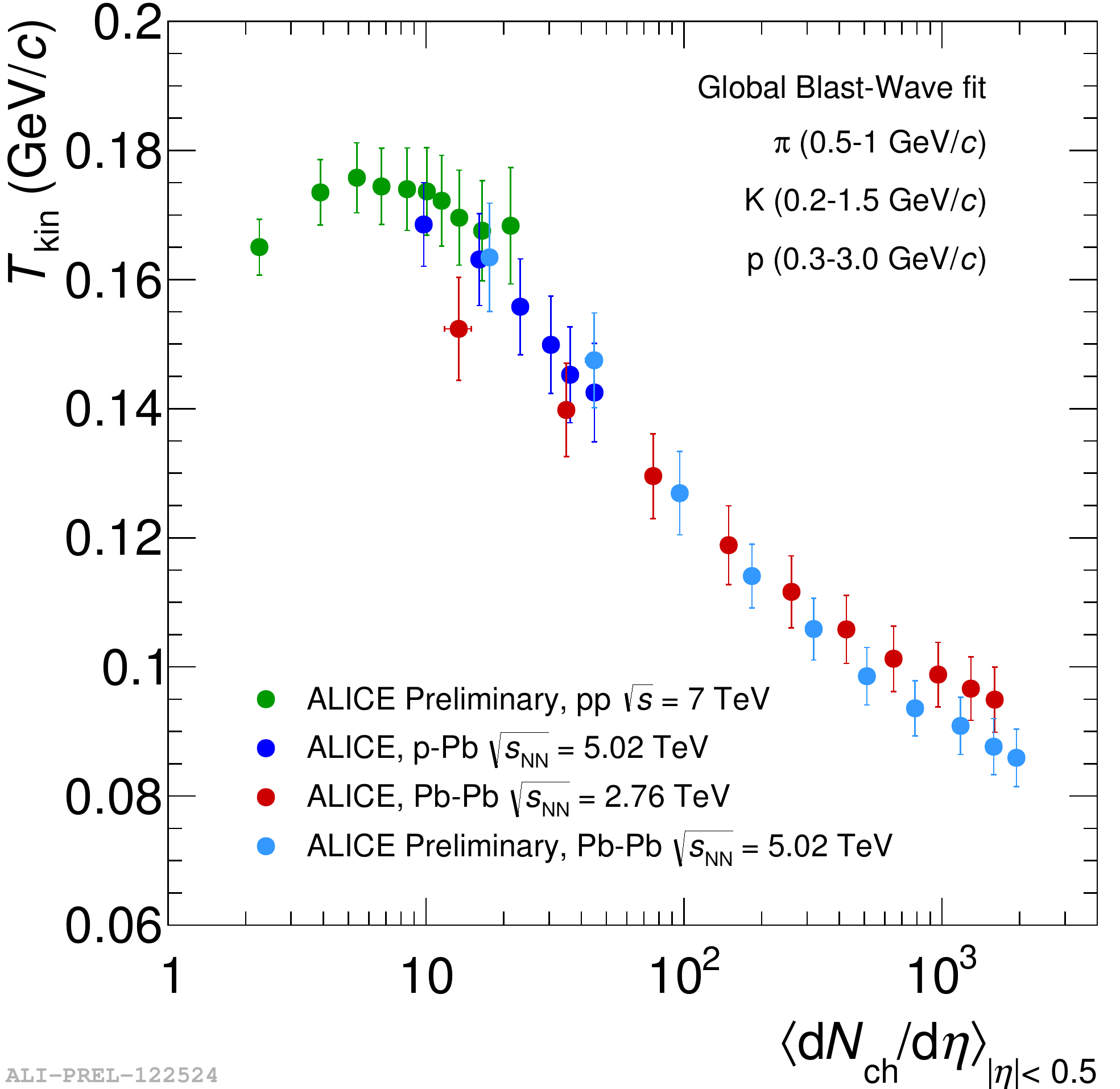}
  \includegraphics[width=0.45\textwidth]{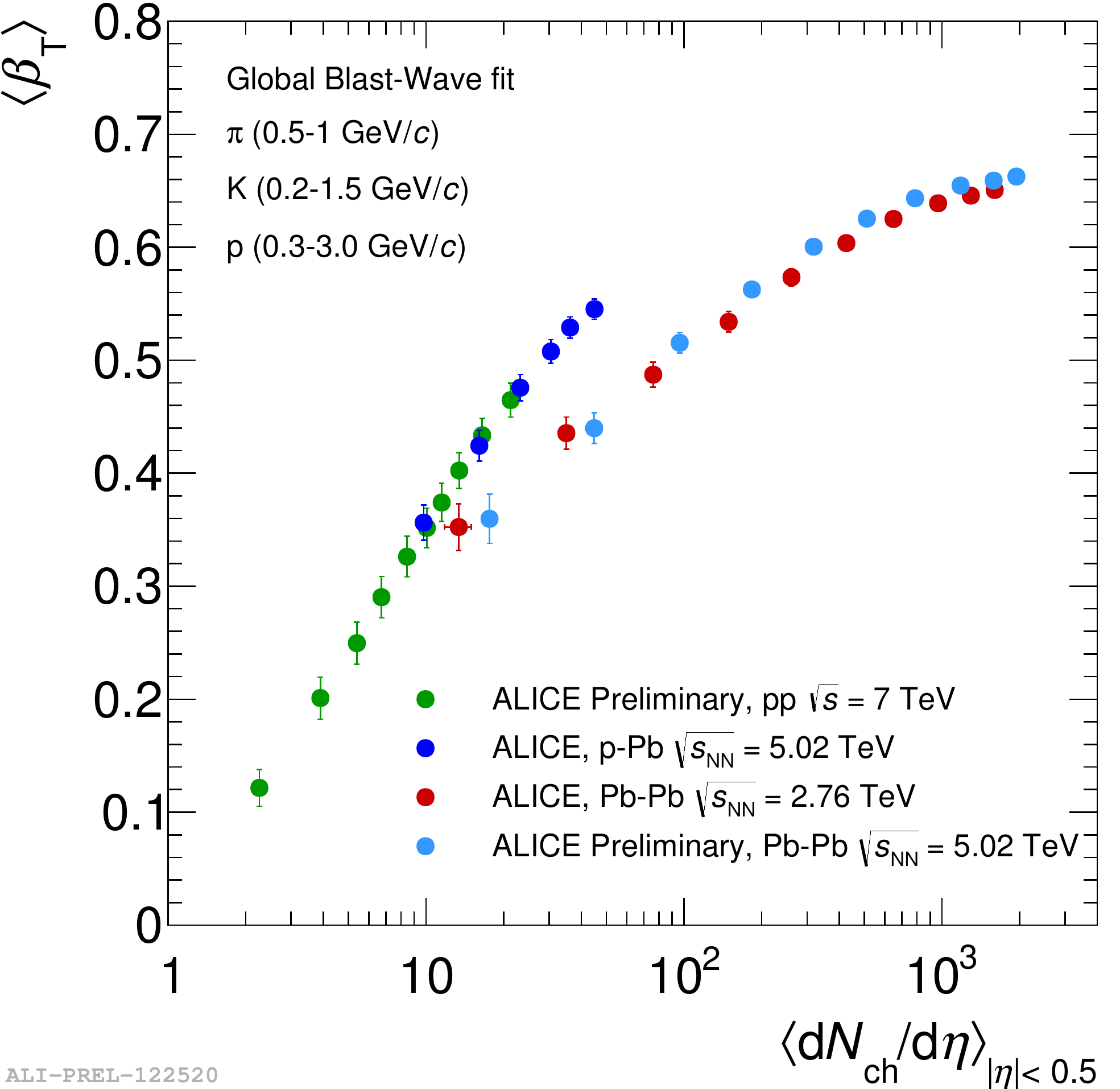}
  \caption{Multiplicity dependence of the \Tkin and \betaT Blast-Wave~\cite{Schnedermann:1993ws} parameters fitted to pion, kaon and proton \pt spectra measured by ALICE in pp, \pPb and \PbPb collisions.}
  \label{fig:collective:spectra:bw}
\end{figure}

\subsection{Identified hadron spectra}\label{sec:collective:spectra}

The \pt distributions of identified hadrons are another tool to probe the collective behaviour of particle production. If collective radial flow develops, this would result in a characteristic dependence of the shape of the transverse momentum distribution on the particle mass.
The \pt distributions in pp and \pPb collisions show a clear evolution, becoming harder as the multiplicity increases, with a change which is most pronounced for protons and lambdas. The stronger multiplicity dependence of the spectral shapes of heavier particles is evident when looking at the \pt-dependent ratios, such as the \sLambda/\pKzero ratio measured by the CMS Collaboration~\cite{Khachatryan:2016yru}, shown in Figure~\ref{fig:collective:spectra:cms}. The ratios in pp and \pPb collisions show a significant enhancement at intermediate \pt ($\sim$3~\GeVc), quantitatively similar to that measured in \PbPb collisions at similar multiplicity.
Results from the ALICE Collaboration show that the modification of the \pt-depentent ratios of \sProton/\sPi and of \sLambda/\pKzero follows a rather smooth evolution at low and mid \pt across different systems, pointing towards a universal soft mechanism driven by final state \dnchdeta~\cite{Abelev:2013haa}. The high-\pt part of the ratio, which is dominated by hard fragmentation, is unchanged.

The Blast-Wave model~\cite{Schnedermann:1993ws} is a useful tool to characterize the spectral shapes of identified hadrons and test data against the radial flow picture. A simultaneous fit to pions, kaons and protons is performed by the ALICE Collaboration following the approach discussed in~\cite{Abelev:2013haa}. The best fit describes the data with good accurancy in the low-\pt domain, stressing the consistency of particle production from a thermal source expanding with a common transverse velocity. The results of the fit to pp, \pPb and \PbPb data are shown in Figure~\ref{fig:collective:spectra:bw} as a function of \dnchdeta.
In the context of the Blast-Wave model, when comparing the parameters of different systems at similar \dnchdeta, \Tkin values are similar, whereas \betaT is larger for small systems. This might be an indication of a larger radial flow in small systems as consequence of stronger pressure gradients.
The ALICE results also show that both \Tkin and \betaT parameters obtained from pp and \pPb fits are consistent for similar \dnchdeta. One has to acknowledge that a different conclusion emerges from similar fits reported by CMS~\cite{Khachatryan:2016yru}, where pp collisions appear as more explosive than \pPb collisions, with a larger \betaT for similar \dnchdeta.

%% file: strangeness.tex
\section{Strangeness production}\label{sec:strangeness}
The study of the production of strange hadrons in high-energy hadronic interactions provides an important means to investigate the properties of QCD.
Unlike up ($u$) and down ($d$) quarks, which form ordinary matter, strange ($s$) quarks are not present as valence quarks in the initial state.
Yet they are sufficiently light to be abundantly created by non-perturbative (soft) processes in the course of the collisions.
String fragmentation, as an example of non-perturbative QCD process, dominates the production of strange hadrons at low \pt.
As a matter of fact, given that the mass of the strange quark is larger than the one of up and down quarks, production of strange hadrons in fragmentation is suppressed relative to hadrons containing only light quarks.
\input{enhancement}
\input{canonical}

%% file: enhancement.tex
\begin{figure}[t]
  \begin{minipage}[b]{.52\textwidth}
    \centering
    \includegraphics[width=0.9\textwidth]{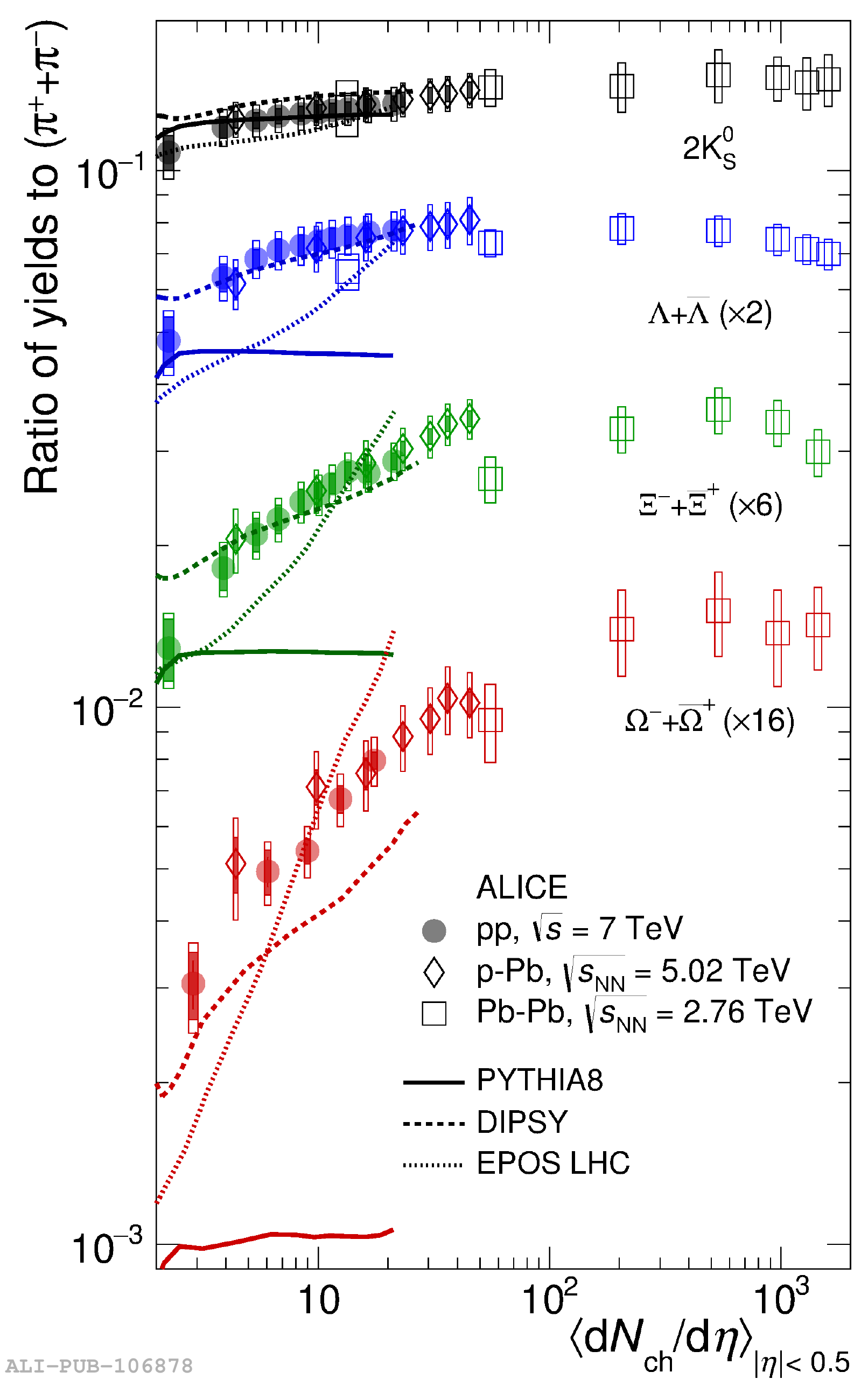}
  \end{minipage}
  \hfill
  \begin{minipage}[b]{.48\textwidth}
    \centering
    \includegraphics[width=0.9\textwidth]{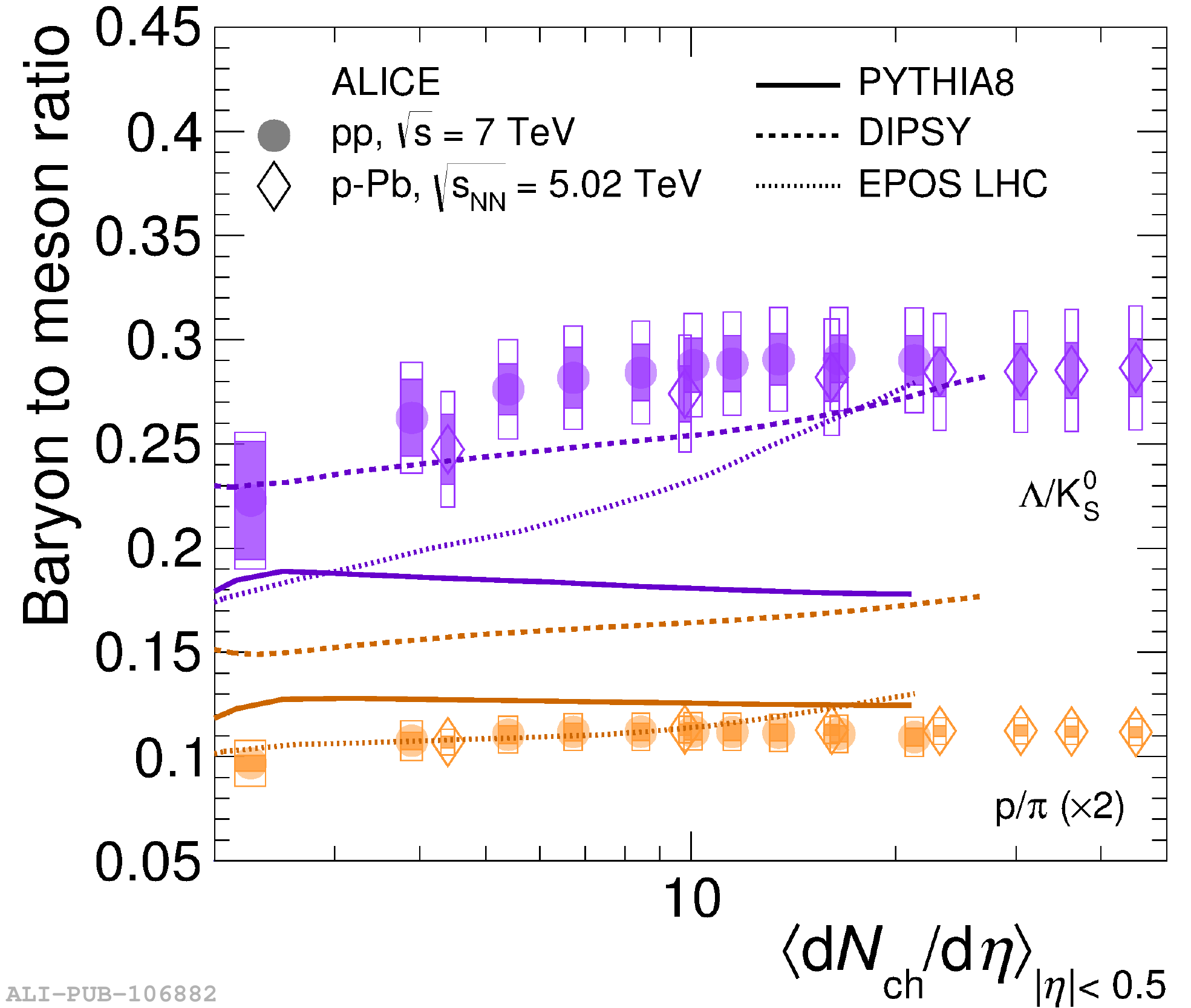}
    \includegraphics[width=0.9\textwidth]{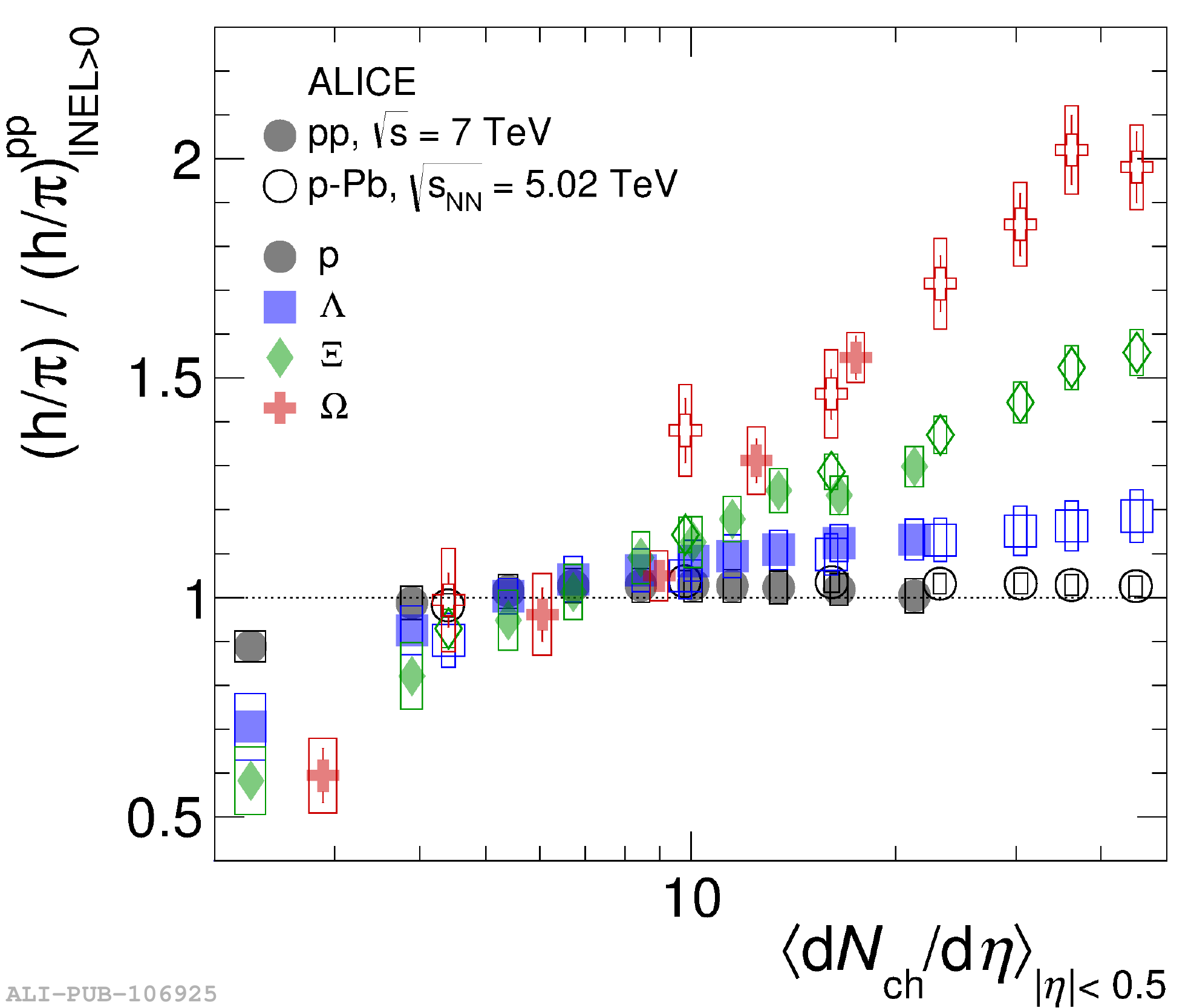}
  \end{minipage}
  \caption{(left) \pt-integrated yield ratios of strange and multi-strange hadrons to pions as a function of \dnchdeta measured by ALICE in pp, \pPb and \PbPb collisions~\cite{ALICE:2017jyt,Abelev:2013haa,Adam:2015vsf,ABELEV:2013zaa}.}
  \label{fig:strangeness:enhancement:ratios}
  \vspace{-5mm}
  \caption{(top right) Particle yield ratios \sLambda/\pKzero and \sPr/\sPi as a function of \dnchdeta measured by ALICE in pp and \pPb collisions. (bottom right) Particle yield ratios to pions normalised to the values measured in the inclusive pp sample.}
  \label{fig:strangeness:enhancement:bmratios}
\end{figure}

\subsection{Strangeness enhancement}\label{sec:strangeness:enhancement}
An enhanced production of strange hadrons was one of the earliest proposed indicators for the formation of a Quark-Gluon Plasma (QGP) state~\cite{Rafelski:1982pu,Koch:1982ij,Koch:1986ud}, as higher rates for strange quark production are expected in a highly-excited state of QCD matter.
This strangeness enhancement is expected to be more pronounced for multi-strange baryons, and was indeed observed in collisions of heavy nuclei at the SPS, RHIC and LHC~\cite{Andersen:1998vu,Andersen:1999ym,Antinori:2004ee,Afanasiev:2002he,Anticic:2003ux,Adams:2003fy,Adams:2006ke,Abelev:2007xp,Antinori:2010jm,ABELEV:2013zaa}. 

The ALICE Collaboration has recently reported on the multiplicity dependence of the production of primary strange and multi-strange hadrons in pp collisions~\cite{ALICE:2017jyt}. 
Similar measurements have been previously performed by ALICE \pPb collisions~\cite{Abelev:2013haa,Adam:2015vsf} and in \PbPb collisions~\cite{Abelev:2013xaa,ABELEV:2013zaa}. Figure~\ref{fig:strangeness:enhancement:ratios} shows the measured ratios of the \pt-integrated yields of \pKzero, \sLambda, \sXi and \sOmega to the pion yield as a function of \dnchdeta in pp collisions~\cite{ALICE:2017jyt} compared to \pPb and \PbPb results at the LHC~\cite{Abelev:2013haa,Adam:2015vsf,ABELEV:2013zaa}.
A significant enhancement of strange to non-strange hadron production is observed with increasing charged-particle multiplicity in pp collisions. The behaviour observed in pp collisions resembles that of \pPb collisions at a slightly lower centre-of-mass energy~\cite{Adam:2015vsf}, both in the values of the ratios and in their evolution with \dnchdeta, suggesting that strangeness production is driven by the event activity of the event rather than by the initial-state collision system or energy.
Figure~\ref{fig:strangeness:enhancement:bmratios} shows that the \pt-integrated yield ratios \sLambda/\pKzero and \sProton/\sPi do not change significantly with multiplicity, demonstrating that the observed enhanced production rates of strange hadrons with respect to pions is not due to the difference in the hadron masses.

The results in Figures~\ref{fig:strangeness:enhancement:ratios} and~\ref{fig:strangeness:enhancement:bmratios} are compared to calculations from Monte Carlo models commonly used for pp collisions at the LHC: PYTHIA8~\cite{Sjostrand:2007gs}, EPOS LHC~\cite{Pierog:2013ria} and DIPSY~\cite{Flensburg:2011kk,Bierlich:2014xba,Bierlich:2015rha}.
The observation of a multiplicity-dependent enhancement of the production of strange hadrons along with the constant production of protons relative to pions cannot be simultaneously reproduced by any of the Monte Carlo event generators commonly used at the LHC.

%% file: canonical.tex
\begin{figure}[t]
\begin{minipage}[b]{.645\textwidth}
\centering
\includegraphics[width=0.9\textwidth]{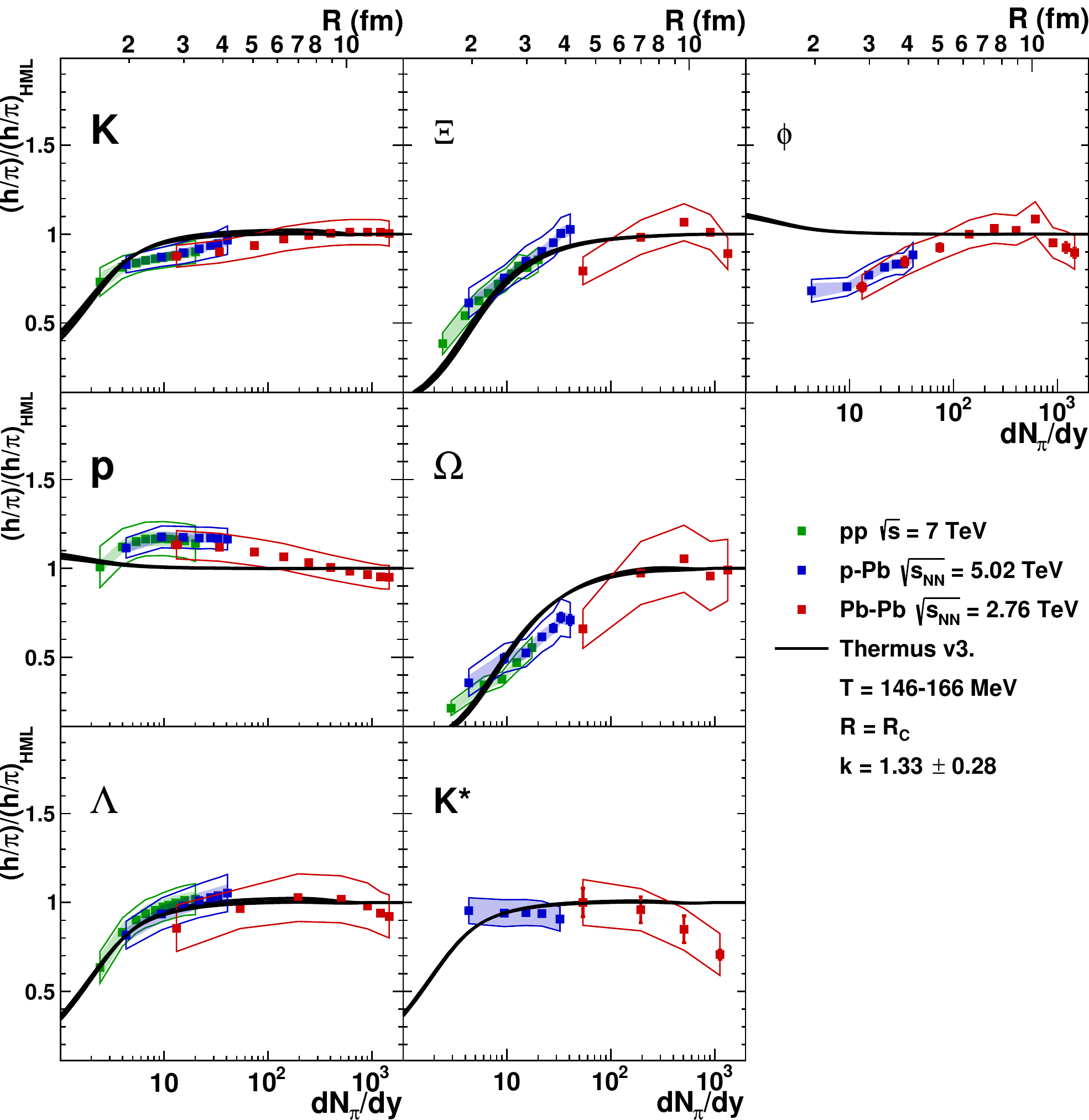}
\end{minipage}
\hfill
\begin{minipage}[b]{.345\textwidth}
\centering
\includegraphics[width=0.9\textwidth]{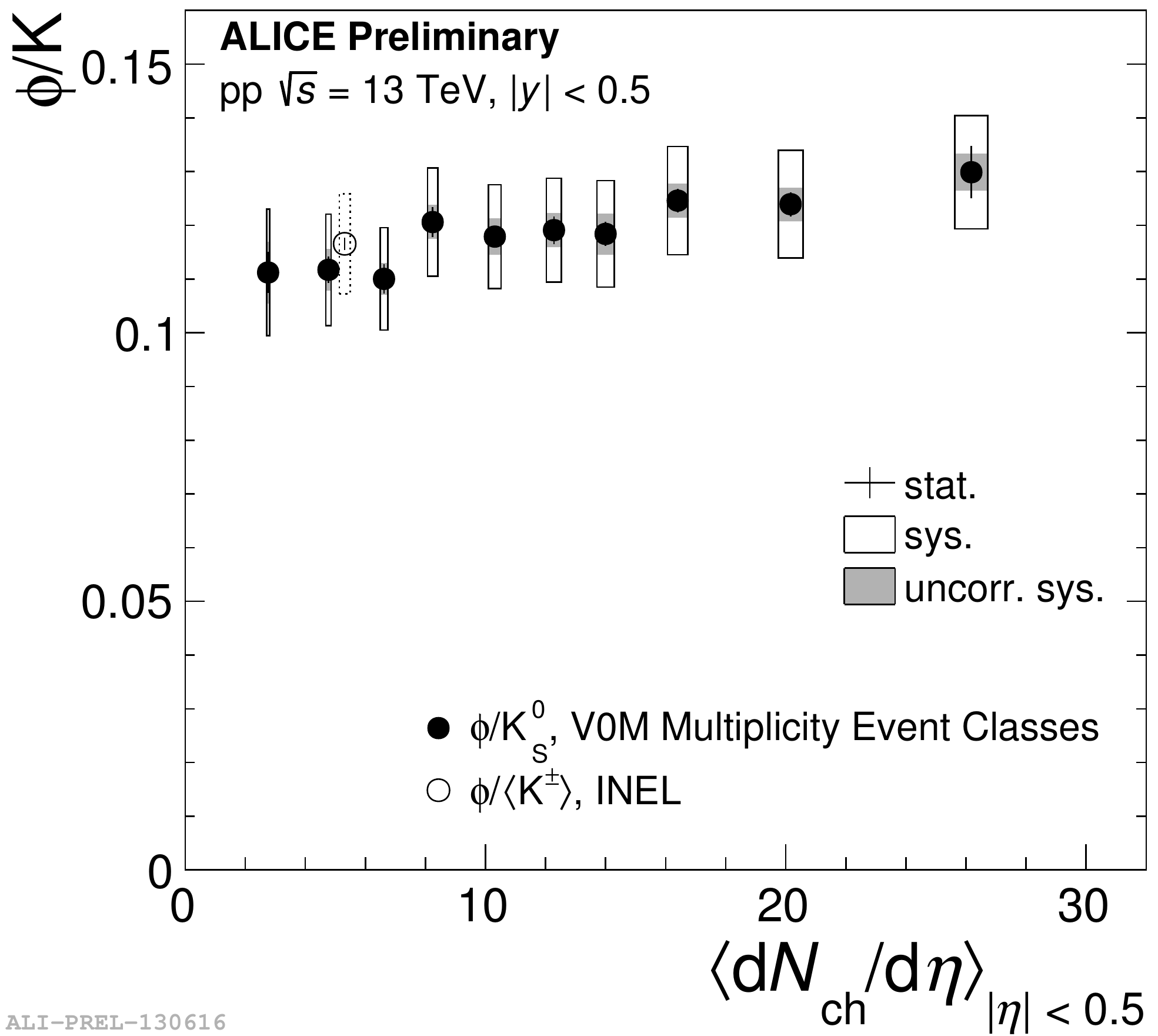}
\includegraphics[width=0.9\textwidth]{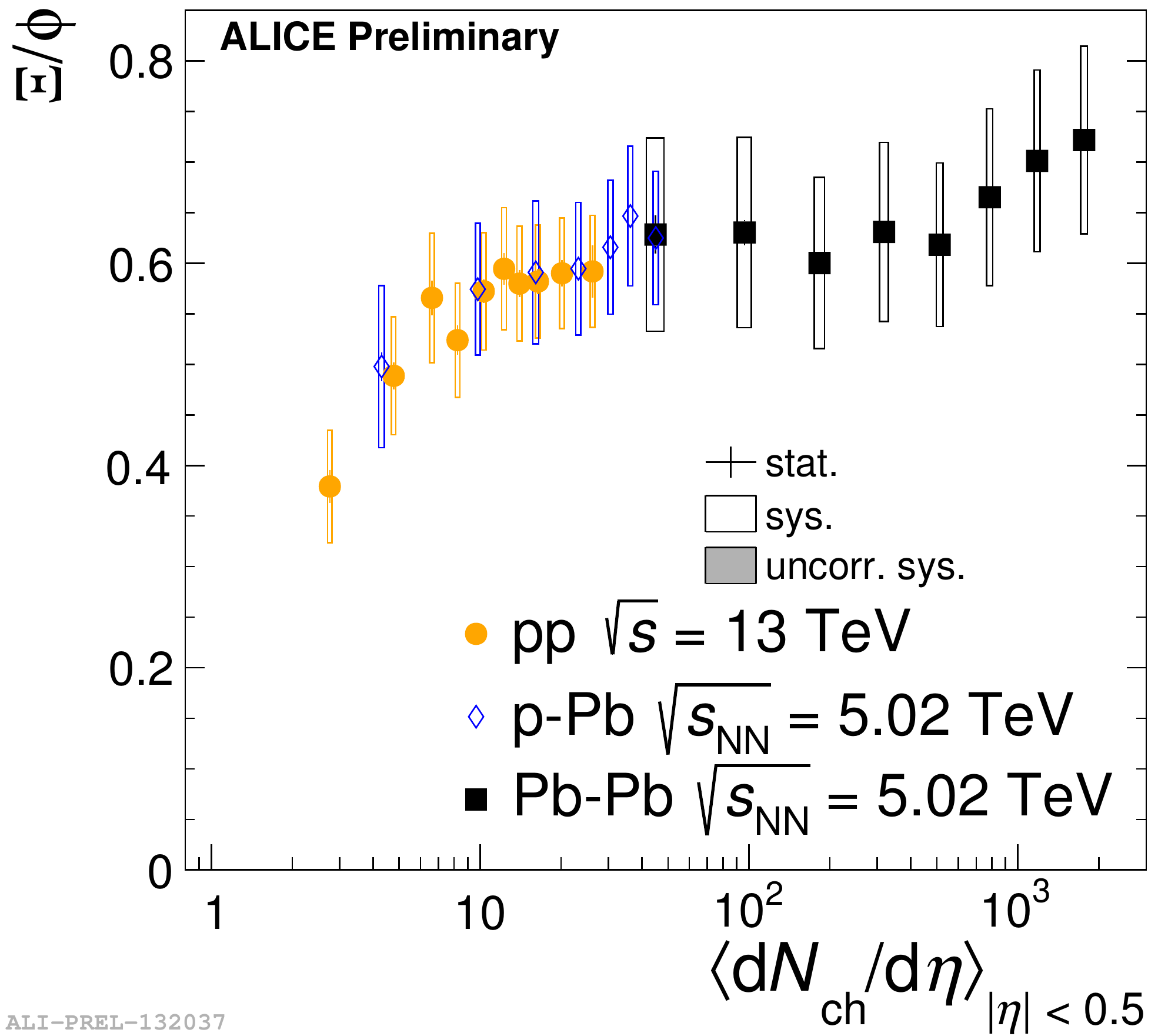}
\end{minipage}
\caption{(left) Ratios of several particle species measured by ALICE as a function of the midrapidity pion yields for pp, \pPb and \PbPb collisions compared to the THERMUS strangeness canonical suppression model prediction~\cite{Vislavicius:2016rwi}.}
\label{fig:strangeness:canonical:suppression}
\vspace{-5mm}
\caption{(top right) \pt-integrated $\phi$/K ratio in pp collisions as a function of \dnchdeta. (bottom right) \pt-integrated $\Xi$/$\phi$ ratio in pp, \pPb~\cite{Adam:2015vsf,Adam:2016bpr} and \PbPb collisions as a function of \dnchdeta.}
\label{fig:strangeness:canonical:phi}
\end{figure}

\subsection{Canonical suppression}\label{sec:strangeness:canonical}
The abundances of strange particles in heavy-ion collisions are compatible with those of a hadron gas in thermal and chemical equilibrium and can be described using a grand canonical statistical model~\cite{Cleymans:2006xj,Andronic:2008gu}.
Extensions of the statistical description, like the ones employed in the strangeness canonical suppression~\cite{Redlich:2001kb} and the core-corona superposition~\cite{Becattini:2008yn,Aichelin:2008mi} models, can effectively produce a suppression of strangeness production in small systems.

The authors in~\cite{Vislavicius:2016rwi} have studied the multiplicity dependence of light flavour hadron production at LHC energies in the strangeness canonical suppression picture using THERMUS~\cite{Wheaton:2004vg}. The details can be found therein. As can be observed from Figure~\ref{fig:strangeness:canonical:suppression}, the study shows that the model provides a good description of the experimental data, except for the $\phi$ meson. As a matter of fact, the strangeness canonical suppression model is only sensitive to the strangeness quantum number of the hadron. The observation of an enhancement (suppression) of $\phi$ meson production with increasing (decreasing) multiplicity clearly signals that the particle production rate is not consistent with the one expected from a hadron with zero strangeness quantum number. This can be observed also in Figure~\ref{fig:strangeness:canonical:phi}, which shows the ratio of $\phi$/K and $\Xi$/$\phi$ as a function of \dnchdeta. It is evident that $\phi$ meson (S = 0) production increases faster than K meson (S = 1) production. When compared to the $\Xi$, the $\Xi$/$\phi$  is constant within uncertainties for \dnchdeta $>$ 10. This is an indication that the $\phi$ meson behaviour is closer to the behaviour expected from a hadron formed by two strange quarks. The increase of the relative $\phi$ meson production with \dnchdeta constitues a notable crack in the canonical suppression interpretation of the observed strangeness enhancement in high-multiplicity pp and \pPb collisions.

%% file: conclusions.tex
\section{Conclusions}\label{sec:conclusions}
Several effects, like near-side long-range correlations and mass-dependent hardening of \pt distributions, which in nuclear collisions are typically attributed to the formation of a strongly-interacting collectively-expanding quark-gluon medium, have been observed in high-multiplicity pp and \pPb collisions at the LHC~\cite{Khachatryan:2010gv,CMS:2012qk,Abelev:2012ola,Aad:2012gla,Aad:2013fja,Chatrchyan:2013nka,Abelev:2013haa,ABELEV:2013wsa,Adam:2015vsf,Khachatryan:2016yru,Khachatryan:2016txc}.
An enhanced production of strange particles as a function of the charged-particle multiplicity density (\dnchdeta), originally considered to be another signature of QGP formation in nuclear collisions~\cite{Rafelski:1982pu,Koch:1982ij,Koch:1986ud}, has also been recently observed in pp collisions~\cite{ALICE:2017jyt}.

The study of small collision systems at high multiplicity is undoubtedly of considerable interest. Further studies are essential to assess whether the combination of these observations can be interpreted as signal of the progressive onset of a QGP medium, which starts developing already in small systems.